# Localized high-Q resonances enable electrically defined beam steering in lithium niobate metasurfaces

Jiyong Shim,[1,*] Sahil Dagli,[1] Hamish Carr Delgado,[1] Remi Dado,[2] Chih-Yi Chen,[1] Feng Pan,[1] Isabella He,[1] Sajjad Abdollahramezani,[1] Mark L. Brongersma,[1] and Jennifer A. Dionne[1]

[1]*Department of Materials Science and Engineering, Stanford University, 450 Jane Stanford Way, Stanford, CA 94305, USA*
[2]*Department of Chemistry, Stanford University, 450 Jane Stanford Way, Stanford, CA 94305, USA*
**Corresponding author: jys7911@stanford.edu*

**Abstract**

High-speed beam steering for LiDAR and optical communication requires independent, reconfigurable control of optical phase. We demonstrate a metasurface combining lithium niobate's Pockels effect with high-Q (>1700) silicon guided-mode resonance nanoantennas whose localized resonances allow individual electro-optic control at a 1 μm pitch. Identical arrays steer light to 53.5° or 32.4°, with the angle set by the bias pattern rather than antenna geometry. We measure 18.0% 0th-order modulation depth, 0.09% of incident power in the steered order, and a 940 MHz detected modulation bandwidth. Mode-overlap analysis shows how inter-element coupling limits usable Q, providing design rules for compact, high-speed beam steering.

## 1. Introduction

High-speed optical beam steering is important for light detection and ranging (LiDAR) [1–3], free-space optical communications [4–6], augmented and virtual reality (AR/VR) displays [7,8], and spatial addressing of atomic and quantum systems [9,10]. These applications motivate compact optical devices capable of redirecting light over large angles [11,12] while rapidly reconfiguring the optical wavefront. Metasurfaces offer a promising platform by controlling the amplitude and phase of light through arrays of nanoscale scatterers [13,14], but their response is typically fixed at fabrication. Steering to large angles further requires the phase of each scatterer to be set independently at a pitch close to or below the wavelength.

Reconfigurable metasurfaces have employed liquid crystals [15,16], phase-change materials [17,18], transparent conducting oxides [19–21], and micro-electromechanical systems (MEMS) [22–24] to dynamically control optical wavefronts. Each approach offers distinct tradeoffs among modulation speed, optical loss, phase tunability, and electrical addressability. The Pockels effect provides an attractive alternative, enabling ultrafast, low-loss refractive-index modulation without relying on carrier injection or structural phase transitions. Among Pockels materials such as electro-optic polymers [25–27] and barium titanate [28,29], lithium niobate (LN), in particular, combines a strong electro-optic response, broad optical transparency, and mature thin-film fabrication [30–33], with integrated modulators demonstrating bandwidths exceeding 100 GHz [34,35]. LN metasurfaces have been widely explored [36–41], with some reaching GHz modulation speeds, but most operate as intensity modulators or impose a fixed wavefront, whereas beam steering requires an electrically reconfigurable phase profile. The small electrically induced index change and short optical interaction length of a metasurface limit the attainable phase modulation. Resonances with a high quality factor (Q) can amplify this response, but they also couple neighboring resonant elements, making it challenging to maintain the spatial localization needed for independent, subwavelength-scale control.

Recent advances have begun to overcome these limitations. Our previous computational work proposed localized, high-Q silicon-on-LN guided-mode resonances for independent electro-optic phase control [42], followed by experimental demonstration of GHz-speed modulation of a fixed optical wavefront [36]. More recently, electro-optically induced diffraction was demonstrated using high-Q silicon-on-LN resonators in a transmissive metasurface [43]. Building on these works, a remaining goal is to electrically define the steering angle itself while maintaining large deflection angles and high-speed operation. Achieving this requires understanding how optical confinement, inter-element coupling, and electrical bandwidth can be engineered together to enable independent control of high-Q resonators at subwavelength scales.

Here, we address these challenges using individually addressable silicon (Si) nanoantennas supporting localized high-Q guided-mode resonances (GMRs) evanescently coupled to thin-film LN (Fig. 1). Each nanoantenna comprises a notched Si waveguide whose periodic perturbations couple guided light to free-space radiation [44–46], while the notch geometry controls radiative leakage and resonance Q [47,48]. The optical mode is predominantly confined within each high-index Si nanoantenna, while partial etching of the underlying LN suppresses coupling between neighboring resonators, enabling localized resonances with Q exceeding 1700 at a 1 μm pitch. An applied voltage shifts each resonance through the Pockels effect, converting small refractive-index changes into local optical phase modulation [42,49]. Temporal coupled-mode theory identifies the conditions for a full $2\pi$ resonant phase excursion, and a mode-overlap analysis shows how inter-element coupling limits the usable Q. Using identical nanoantenna geometries with distinct electrode configurations, we experimentally demonstrate electrically defined beam steering to 53.5° and 32.4°, with more than 18 dB of angular contrast and directly measured power in the steered diffraction orders. We further demonstrate electro-optic modulation bandwidths up to 940 MHz in the detected second-harmonic response and establish their scaling with metasurface dimensions. Together, these results advance electrically controlled resonant metasurfaces beyond fixed diffraction configurations, providing quantitative design principles for integrating wide-angle beam steering, localized high-Q resonances, and high-speed electrical control.

## 2. Results

### 2.1 Design of the GMR nanoantenna

Fig. 1(a) shows the metasurface with notched Si GMR nanoantennas on a co-patterned LN film, with Ag electrodes shared between neighboring antennas. Because the periodic notches let the bound guided mode radiate into free space, it appears as a resonance whose quality factor is set by the notch size [47,48].

With this design principle in mind, we implement the platform as a linear array of localized GMR nanoantennas. The GMR nanoantennas are 300 nm tall and placed atop a 300 nm thick LN film that is etched to a residual thickness of 120 nm between antennas. Numerical simulations indicate that 6.7% of the modal energy resides within the underlying LN layer.

We fabricate this design following standard nanofabrication procedures (see Methods), with the nanoantennas patterned via e-beam lithography to ensure high structural fidelity. Fig. 1(d) shows a tilted SEM image of the fabricated $200 \times 200$ μm² metasurface (inset: optical micrograph), in which each GMR nanoantenna sits between two electrodes, each shared with the neighboring antenna. Fig. 1(e) shows an SEM image with FIB cross-section inset confirming the realized geometry. Fig. 1(f) shows the simulated GMR optical field profile for a notch size of $w = d = 80$ nm.

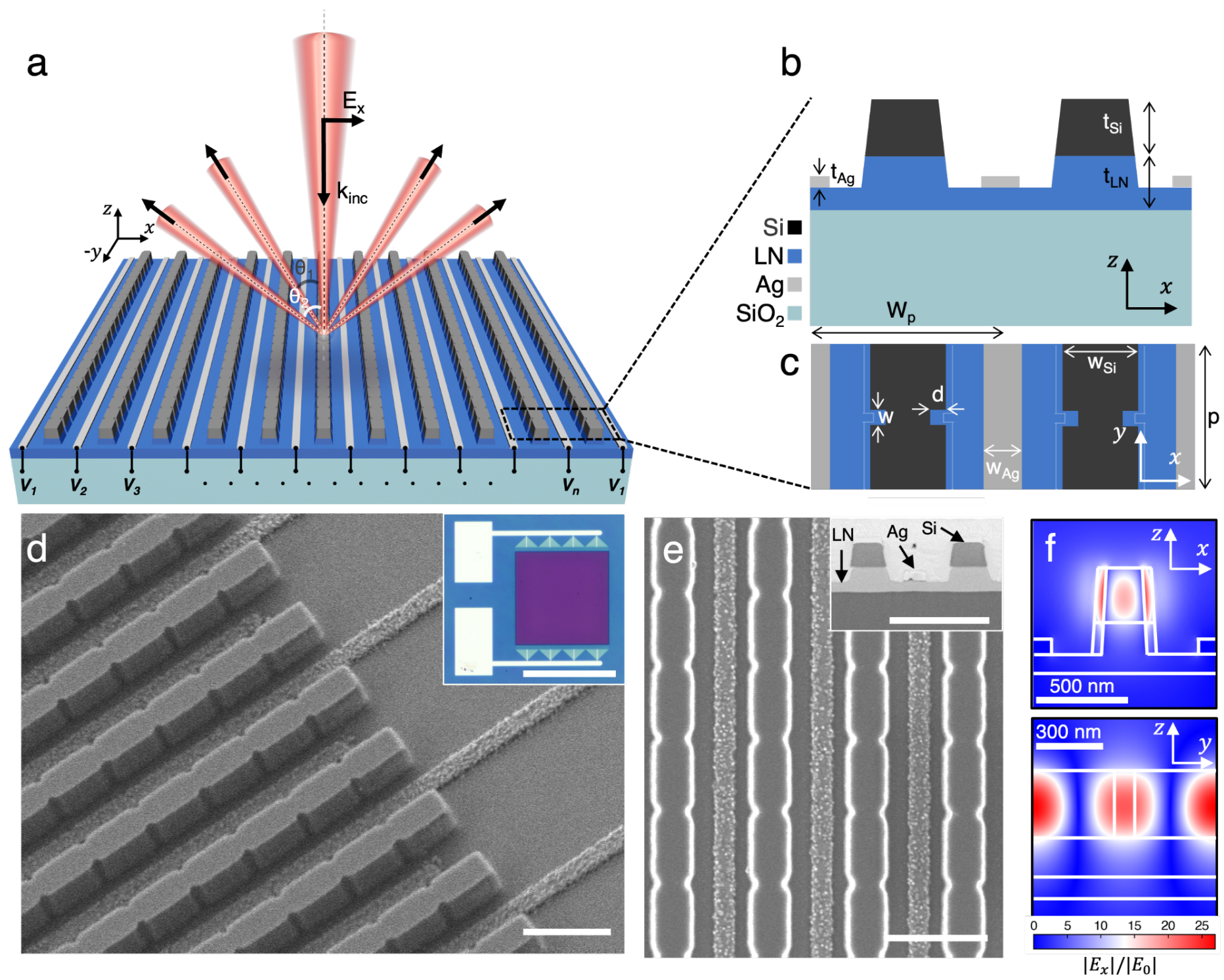

**Fig. 1. Electro-optic beam steering metasurface on a lithium niobate thin film. (a)** Schematic of the device. An x-polarized incident beam ($k_{inc}$) is diffracted into steered orders via electro-optically induced phase modulation. **(b,c)** Unit-cell schematics in the xz (b) and xy (c) planes, showing the silicon nanoantenna, LN thin film, and silver electrodes on a fused silica substrate, with geometric parameters p = 740 nm, w = 80 nm, d = 80 nm, $w_{Si}$ = 380 nm, $t_{Si}$ = 300 nm, $t_{LN}$ = 300 nm (120 nm residual between antennas), $t_{Ag}$ = 70 nm, $w_{Ag}$ = 200 nm, and $w_p$ = 1000 nm. **(d)** 25° tilted scanning electron micrograph (SEM) of the device (scale bar: 1 µm), with optical micrograph inset (scale bar: 200 µm). **(e)** SEM and (inset) focused ion beam cross-section of a fabricated device, confirming the realized geometry (scale bars: 1 µm). **(f)** Simulated near-field enhancement of the guided mode resonance in the silicon nanoantenna for a notch size of w = d = 80 nm at the resonant wavelength of 1606.6 nm.

## 2.2 Resonant phase response and inter-element coupling

For an array idealized as infinitely periodic and supporting a resonance such as a quasi-bound state in the continuum (qBIC) or a guided mode resonance, the reflected response is captured by the reflection coefficient [49–51] (see Supplementary Section 1):

$$r(\omega) = r' + \frac{2\gamma_r}{[i(\omega - \omega_0) + \Gamma_{tot}]} \quad (1)$$

$$\Gamma_{tot} = \gamma_{abs} + \sum_m \gamma_m \quad (2)$$

where $r'$ is the phase-normalized background (non-resonant) reflection coefficient, $\gamma_r$ is the radiative coupling rate into the reflection channel, and $\Gamma_{tot}$ is the total decay rate, summing the absorptive rate $\gamma_{abs}$ and the radiative rate into every open channel m, including reflection and transmission. As ω is swept through resonance, $r(\omega)$ traces a circle in the complex plane with radius $\gamma_r/\Gamma_{tot}$, centered at $r' + \gamma_r/\Gamma_{tot}$. A full 2π phase span is achieved only when this circle encloses the origin. For a two-port system, this is equivalent to $\gamma_r > \gamma_t + |r'|^2\gamma_{abs}$, where $\gamma_t$ is the radiative rate into the transmission channel (Supplementary Section 1). For efficient beam steering, it is preferable for the resonance to exhibit this full 2π phase change while maintaining a comparatively high reflectance.

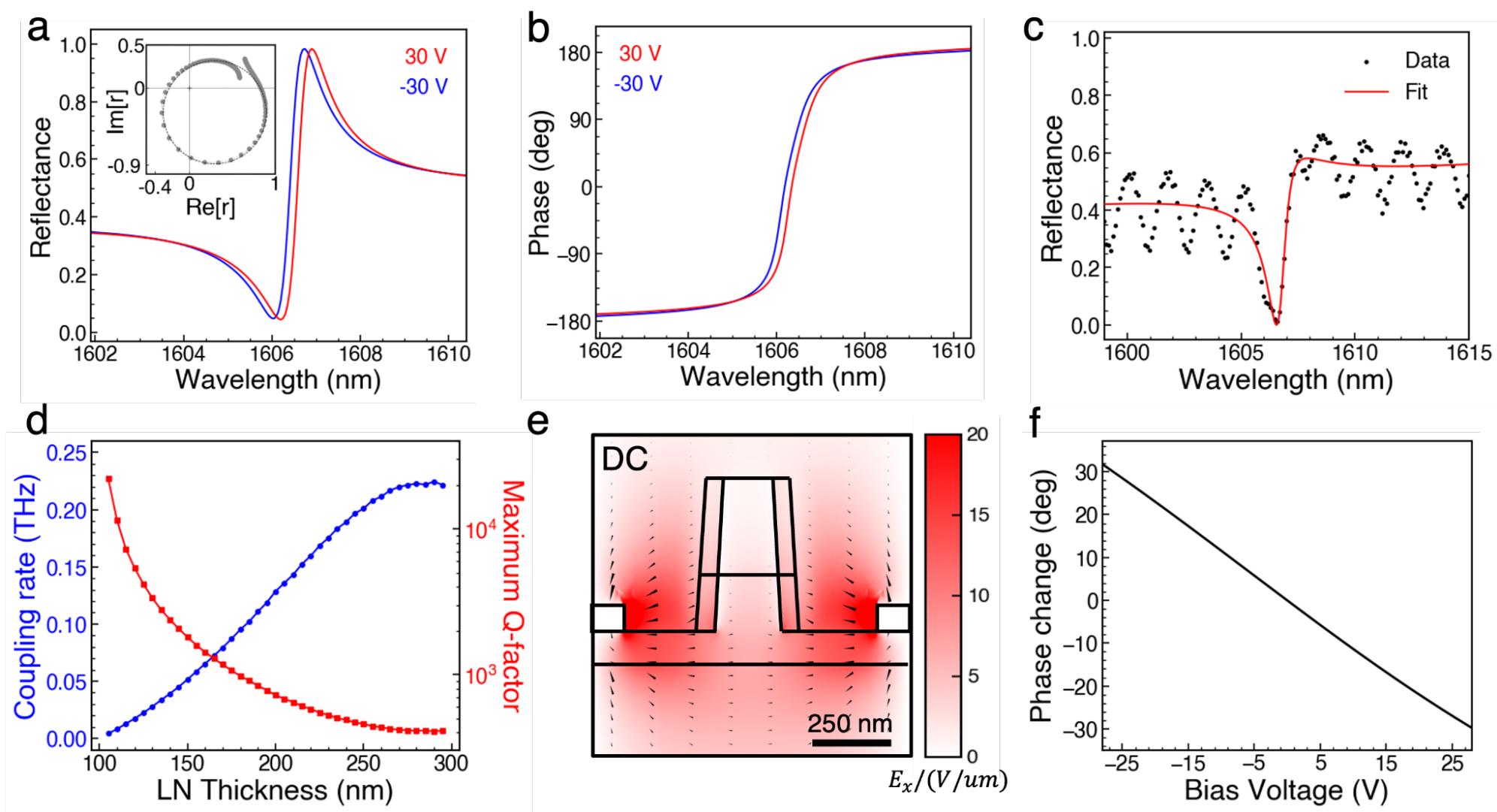

**Fig. 2. Electro-optic phase modulation of the guided mode resonance within a unit cell. (a,b)** Voltage-induced shift of the resonant wavelength, arising from the change in refractive index of the LN, modulates both the intensity (a) and phase (b) of the reflected light. Inset in (a) shows a trajectory of the complex reflection coefficient *r* as the wavelength is swept through the resonance. The circle encloses the origin, indicating a full 2π phase span. **(c)** Measured reflection spectrum of a fabricated device. A Fano-Lorentzian fit yields a quality factor of 1745 at a resonant wavelength of 1606.8 nm. **(d)** Dependence of inter-resonator coupling rate on the residual LN thickness, with the corresponding upper bound on quality factor required to remain in the weakly coupled regime. **(e)** Simulated electric field distribution generated by an applied bias of 10 V across the silver electrodes. **(f)** Voltage-dependent phase modulation at the wavelength of maximum phase sensitivity.

Our design satisfies this condition in two steps. First, we make the metasurface a two-port system by keeping the resonator pitch subwavelength in the substrate ($w_p$ = 1000 nm < $\lambda/n_{sub}$ ≈ 1103 nm, where λ = 1600 nm and $n_{sub}$ = 1.45 for fused silica). Only the 0th-order reflection and transmission channels then remain radiative, maximizing the achievable circle radius $\gamma_r/\Gamma_{tot}$. Second, the substrate and the sloped sidewalls break the mirror symmetry in z, which, together with the Fabry–Perot like background reflection of the unnotched ridge, allows $\gamma_r$ to differ from $\gamma_t$. We choose the silicon ridge width and thickness ($w_{Si}$ = 380 nm, $t_{Si}$ = 300 nm) so that $\gamma_r$ exceeds $\gamma_t$, giving $\gamma_r/\Gamma_{tot} = 0.58$. Together, these achieve a full 2π phase change at resonance with a simulated quality factor of 2020 at 1606.6 nm, as shown in Figs. 2(a) and 2(b). We confirm this resonant response experimentally in Fig. 2(c). Fitting the measured spectrum to a Fano-type resonance lineshape yields a quality factor of 1745 at 1606.8 nm.

Because this reflection coefficient is a property of the repeated resonators, realizing an independent phase response from each nanoantenna also requires that coupling between neighboring elements remain weak. However, because these are high-Q resonators placed in proximity, evanescent coupling between adjacent nanoantennas can hybridize their modes and result in a delocalized response (see Supplementary Section 3). The coupling rate between neighboring resonators is set by the degree of mode overlap between them, and in the Si/LN platform the LN, of higher index than the air above, provides the dominant coupling channel, since the mode's evanescent tail extends further into the LN than into air. The coupling rate is therefore highly sensitive to the residual LN thickness connecting adjacent resonators [Fig. 2(d)]. Localized, independently addressable resonances require the inter-antenna coupling rate to remain below the total damping rate, $\kappa < \Gamma_{tot} = \omega_0/2Q$, which places an upper bound $Q < \omega_0/2\kappa$ on the usable quality factor. The device of Fig. 2(c) has Q = 1745 at 1606.8 nm, corresponding to $\Gamma_{tot}/2\pi$ = 53.5 GHz. A residual LN thickness of 120 nm yields $\kappa/2\pi$ = 17.4 GHz. Thus, the experimentally realized array satisfies $\kappa/\Gamma_{tot} \approx 0.33$, placing the resonators well within the weakly coupled regime required for localized per-element electro-optic control.

### 2.3 Electro-optic control of the local phase

An applied bias sets the electric field in the LN, which shifts the resonance through the Pockels effect and thereby tunes the local phase at a fixed operating wavelength. Fig. 2(e) shows the simulated electric field distribution within the unit cell for a 10 V bias applied across the adjacent Ag electrodes. This electric field, E, modulates the permittivity, $\varepsilon$, of the underlying LN through the Pockels effect

$$\Delta\left(\frac{1}{\varepsilon}\right)_i = \sum_j r_{ij} E_j \quad (3)$$

producing the permittivity change shown in Fig. S7. We use x-cut LN, oriented such that the applied field aligns with the crystallographic axis associated with the largest Pockels coefficient, thereby maximizing the electro-optic response ($r_{33}$ = 31 pm/V, compared to $r_{13}$ = 9 pm/V) [33]. The resulting permittivity change shifts the effective mode index of the GMR and, in turn, its resonant wavelength and associated phase response [Figs. 2(a) and 2(b)].

Because the resonance wavelength is voltage-tunable, the local phase of each nanoantenna can also be tuned at a fixed operating wavelength, $\lambda_0$. For the device we design, a bias sweep of ±25 V yields a phase change of π/3 [Fig. 2(f)]. The resonance-circle framework shows that the full 2π span is supported at this Q, but traversing it electro-optically requires the bias to shift the resonance by an appreciable fraction of its linewidth, which calls for either a higher Q or a larger applied voltage. Because the applied field is limited by the coercive field of the LN, increasing Q offers the more practical route to a larger accessible span (see Supplementary Section 2).

Each GMR nanoantenna therefore provides a voltage-controlled local phase, allowing us to discretize a spatial phase profile across the array into a periodic grating. The resulting reconfigurable supercell periodicity P diffracts normally incident light into orders satisfying Eq. (4) below [52].

$$\theta = \sin^{-1}\left(\frac{\lambda_0}{P}\right) \quad (4)$$

### 2.4 Push-pull beam steering

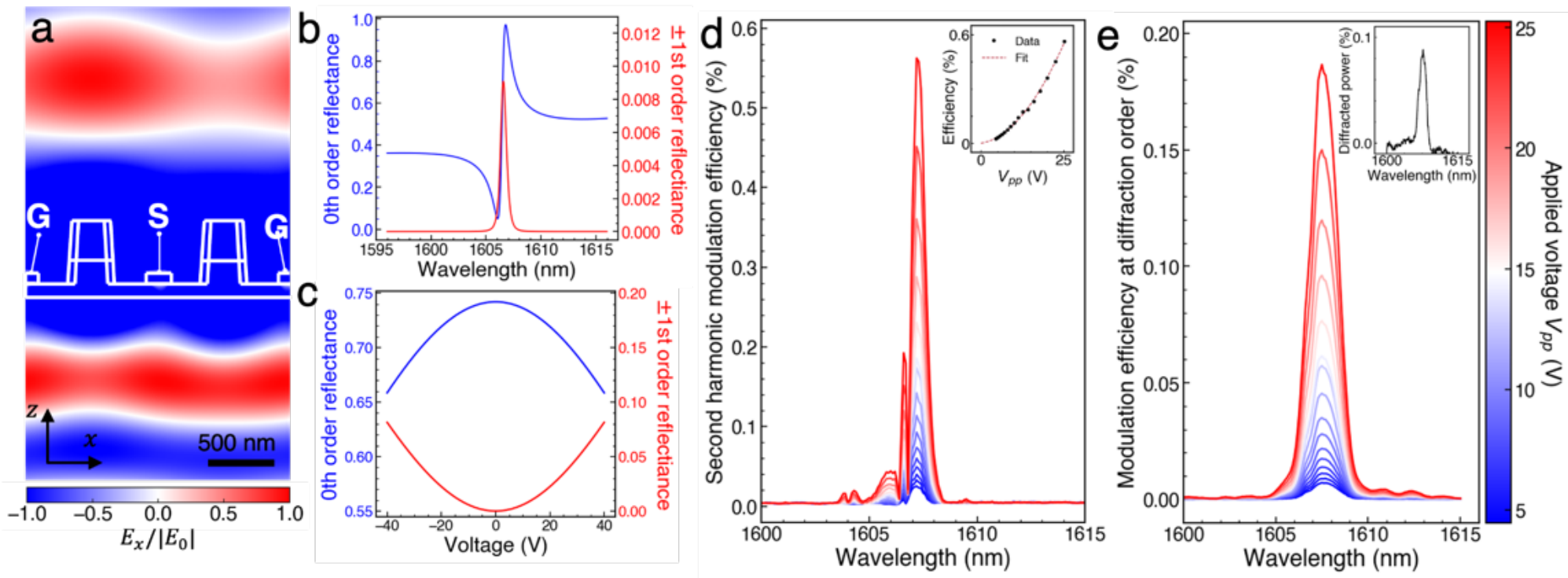

**Fig. 3. Electro-optic beam steering with a push-pull bias design. (a)** Simulated reflected field distortion at an applied bias of 25 V, showing how the push–pull drive (implemented via a ground–signal–ground, G-S-G, biasing scheme) imprints a periodic phase modulation across the array that diffracts light into the ±1 diffraction orders. **(b)** Simulated 0th and ±1st order reflectance spectra at $V_{pp}$ = 25 V, with approximately 0.9% of the incident power diffracted. **(c)** Simulated diffraction efficiency as a function of applied voltage, showing a quadratic dependence that is symmetric with respect to voltage polarity. **(d)** Measured second harmonic modulation efficiency of the 0th order reflected signal, reaching 0.56% at $V_{pp}$ = 25.2 V. The inset shows the quadratic voltage dependence. **(e)** Direct measurement of the diffracted beam, showing 0.19% modulation efficiency at the first diffraction order angle, with the inset showing that 0.09% of the incident optical power is delivered, time-averaged, to the first diffraction order.

To electro-optically steer the light, we adopt a push-pull electrode configuration, in which the applied bias alternates in direction between adjacent unit cells. As shown in Fig. 3(a), this alternating bias imposes an antisymmetric phase modulation across the two-element supercell ($P = 2w_p$ = 2000 nm), diffracting a fraction of the incident power into higher diffraction orders. Because the push-pull profile is symmetric, light is diffracted equally into the +1 and −1 orders, and we detect one of them. Simulations predict that a 25 $V_{pp}$ drive diffracts 0.9% of the incident power into the sum of the ±1 orders [Fig. 3(b)]. The diffraction efficiency scales quadratically with applied voltage [Fig. 3(c)], as expected for a phase grating operating in the small-signal regime. For this two-level grating, 1st-order efficiency is maximized at a phase difference of $\pi$ rather than $2\pi$, and a notch size of 60 nm is predicted to reach this optimum (Fig. S4). Because the reflectance change is an even function of the applied voltage, the modulated signal appears predominantly at the second harmonic of the drive frequency (see Supplementary Section 6).

We measure this modulation using a photodetector connected to an electrical spectrum analyzer. Fig. 3(d) shows a measured second-harmonic modulation efficiency of 0.56% in the 0th order reflected signal at 25.2 $V_{pp}$, with the signal scaling quadratically with applied voltage (inset), and a peak modulation depth of 18.0% at 1606.6 nm [Fig. S9(a)]. The measured modulation efficiency is lower than the simulated diffraction efficiency, consistent with the experimentally reduced Q and fabrication-induced nonuniformity, which causes individual nanoantennas to resonate at slightly different wavelengths.

To measure the steered beam itself, we position a second photodetector along the diffracted beam path, at the angle predicted by Eq. (4), $\theta = \sin^{-1}(1606.8\text{ nm}/2000\text{ nm}) = 53.5°$. Fig. 3(e) shows a measured modulation efficiency of 0.19% at the first diffraction order angle, confirming electro-optic modulation of the steered beam. The inset shows that the time-averaged optical power delivered to this diffraction order reaches 0.09% of the incident power, directly confirming optical power transfer into the steered order. The 0th order [Fig. 3(d)] and diffracted beam [Fig. 3(e)] measurements use different collection geometries and therefore show slightly different lineshapes (Supplementary Section 9).

## 2.5 Electrically defined steering angle

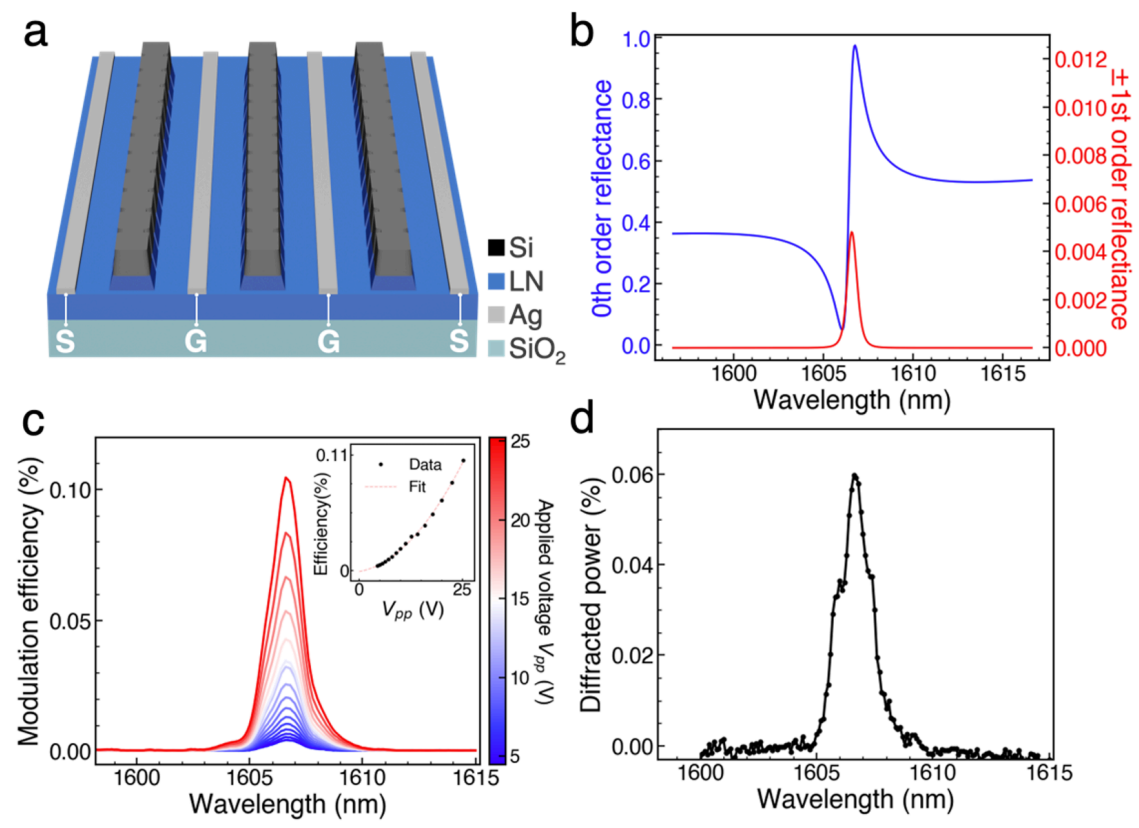

**Fig. 4. Electrically defined steering angle using a different bias pattern. (a)** Schematic of the signal–ground–ground–signal (S-G-G-S) biasing pattern, which defines a three-element supercell. **(b)** Simulated 0th and ±1st order reflectance spectra at $V_{pp} = 25$ V, showing approximately 0.5% of the incident power diffracted to a steering angle of 32.4°. **(c)** Measured modulation spectrum at this diffraction angle, reaching a modulation efficiency of 0.1%. Inset shows the modulation efficiency scaling quadratically with applied voltage. **(d)** Diffracted power measured directly at the reconfigured steering angle, showing that 0.06% of the incident optical power is delivered to the first diffraction order.

The same nanoantenna geometry can support different steering angles when the electrical bias pattern is changed. We demonstrate this principle using a separate sample fabricated with the same nanoantenna geometry but a signal–ground–ground–signal (S-G-G-S) electrode configuration [Fig. 4(a)]. The metasurface exhibits a resonance at 1606.1 nm with a quality factor of 1953 (Fig. S8). The two samples exhibit nearly identical resonance wavelengths (1606.8 and 1606.1 nm), while the effective supercell period changes from 2000 to 3000 nm, resulting in steering angles of 53.5° and 32.4°, respectively. Fig. 4(b) shows that a 25 $V_{pp}$ drive diffracts 0.5% of the incident light into 32.4° in simulation. We measure a second harmonic modulation efficiency of 0.1% directly at this diffraction angle [Fig. 4(c)]. The 0th order signal shows a modulation efficiency of 0.38% (Fig. S8) and a modulation depth of 12.2% [Fig. S9(b)], following the same quadratic dependence on applied voltage established above. Fig. 4(d) confirms that 0.06% of the incident optical power is delivered to this diffracted order via the electro-optic Pockels effect. Measuring both samples at both angles confirms that each bias pattern directs light only to its designed angle, with a contrast of more than 18 dB [Fig. S9(c,d)]. These results demonstrate that the steering angle is determined by the electrically defined diffraction supercell. A single chip could therefore address multiple angles given one contact pad per nanoantenna and a multichannel driver (Fig. S12). The ground-signal probe used here supplies a single bias pattern per sample, so the two configurations were implemented on separate samples fabricated with identical nanoantenna geometries, isolating the bias pattern as the only variable.

## 2.6 High-speed modulation bandwidth

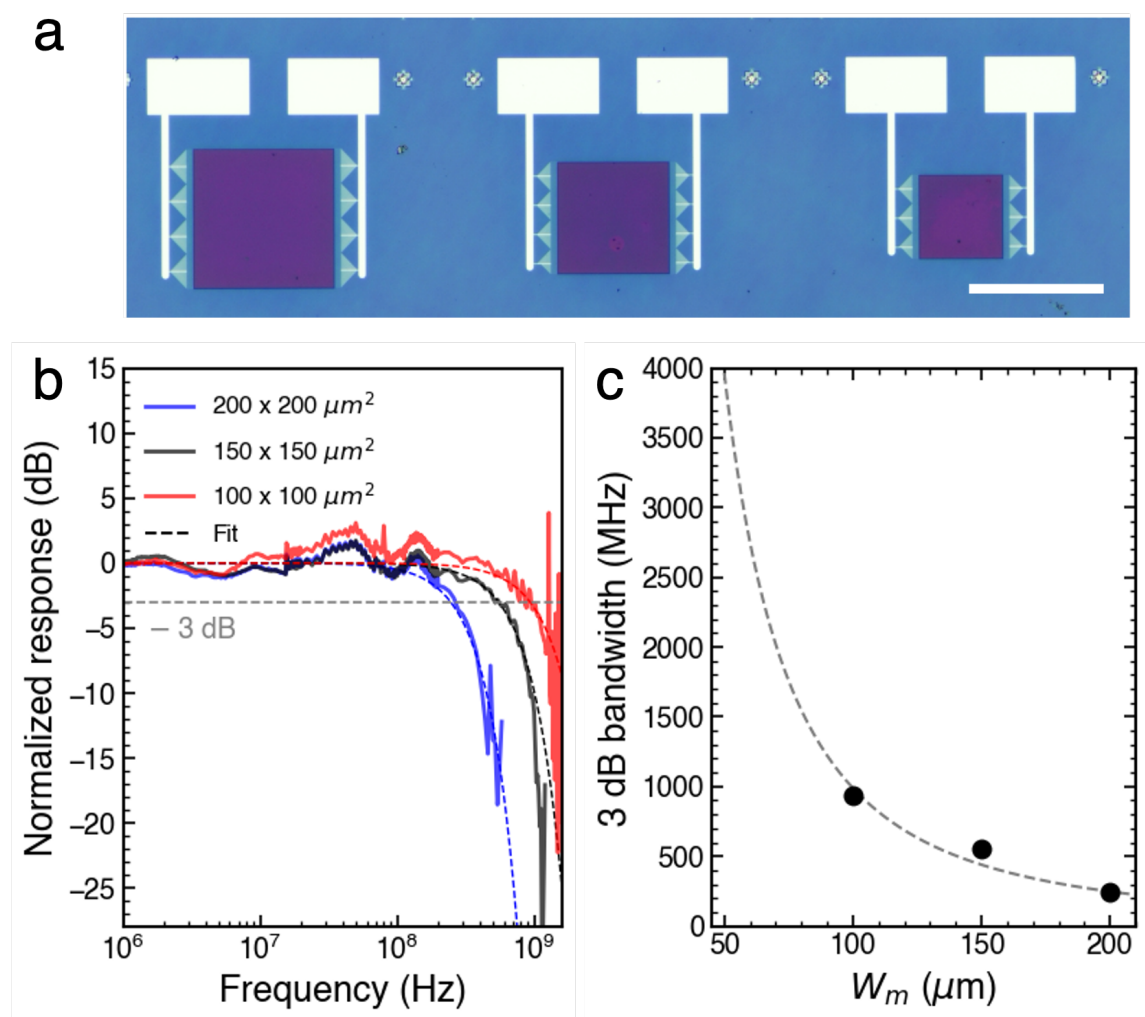

**Fig. 5. Electro-optic modulation bandwidth of the beam-steering metasurface. (a)** Optical microscope images of devices with metasurface widths $W_m$ spanning 100–200 µm (scale bar: 200 µm). **(b)** Modulation response as a function of detected (second harmonic) frequency for each device size in (a), normalized to its value at 1 MHz. **(c)** Extracted 3 dB bandwidth as a function of $W_m$. The dashed line is an inverse-square fit.

Because the electrode capacitance scales with device area, we fabricated devices with three metasurface areas (200 × 200, 150 × 150, and 100 × 100 µm²) to examine how bandwidth scales with device size [Fig. 5(a)]. Each device was driven from 500 kHz to 0.8 GHz while the second harmonic response of the 0th order beam was recorded from 1 MHz to 1.6 GHz. Bandwidths are quoted as the detected modulation frequency, which is twice the drive frequency, and the 3 dB bandwidth was defined as the frequency at which the response first falls 3 dB below its low-frequency baseline [Fig. 5(b)]. Bandwidth increased markedly with decreasing metasurface size, reaching 245 MHz, 552 MHz, and 940 MHz for the 200 × 200, 150 × 150, and 100 × 100 µm² devices, respectively. The bandwidth rises roughly as the inverse square of the metasurface width, as expected if the area-proportional electrode capacitance contributes to the roll-off (Supplementary Section 7 and Table S1). Extrapolation of the measured area scaling projects a bandwidth of approximately 3.6 GHz for a 50 × 50 µm² device [Fig. 5(c)].

## 3. Discussion and conclusion

We have demonstrated an electro-optically controlled metasurface that combines localized high-Q resonances (Q > 1700) with subwavelength-scale electrical addressing to enable wide-angle beam steering. The diffraction supercell is defined by the electrical bias pattern rather than the fabricated nanoantenna geometry. Using identical resonator geometries with distinct bias configurations, we achieve steering angles of 53.5° and 32.4° with more than 18 dB of angular contrast. We further demonstrate electro-optic modulation bandwidths up to 940 MHz in the detected second-harmonic response, with bandwidth increasing as device dimensions decrease. These results establish a route toward high-speed, electrically defined optical wavefront control.

An important insight from this work is that high-Q resonant enhancement and independent electrical addressability impose competing design requirements. Higher Q increases the achievable phase response, but per-element control requires the inter-element coupling rate to remain below the resonance damping rate, which decreases as Q increases. A mode-overlap analysis shows that the residual lithium niobate thickness largely determines this coupling and hence the maximum usable Q. As a second design rule, temporal coupled-mode theory shows that a full 2π phase span requires the resonance to radiate preferentially into the reflection channel, which we achieve by breaking the mirror symmetry in z. These design rules also point toward improvements in steering efficiency, which currently remains limited by the electrically accessible phase shift and fabrication-induced resonance nonuniformity. Optimizing resonator geometry, improving fabrication precision, and refining electrical drive circuitry could increase efficiency while preserving high-speed, localized control.

More broadly, because the optical function is set by the electrical bias pattern rather than the fabricated geometry, a single resonator array can in principle implement different wavefronts. Providing one contact pad per nanoantenna with a multichannel driver would enable dynamic selection of steering angles on a single device, as well as more complex one-dimensional wavefronts such as reconfigurable cylindrical lenses. This approach could enable fast and compact beam steering for LiDAR, free-space optical communication, and addressing linear arrays of atoms or ions. Extending the platform to two dimensions by pixelating each nanoantenna while maintaining high Q [47] would open further applications, such as holographic AR/VR displays.

## 4. Methods

### 4.1 Full-wave simulation of the electro-optic metasurface

Full-wave electromagnetic and electrostatic simulations were performed in COMSOL Multiphysics. The unit cell period was chosen to be subwavelength to suppress higher-order diffraction, and the notch period was set to place the guided mode resonance within the tuning range of the laser. The electro-optic response was evaluated in two steps. The electrostatic field was first solved for a bias applied across the electrodes and converted to an LN index perturbation through the Pockels effect. The perturbed index profile was then used in the full-wave model to extract the resonance shift as a function of applied voltage (Supplementary Section 4).

### 4.2 Device fabrication

Thin-film Si was direct-bonded to an LN-on-insulator (LNOI) substrate. The silicon-on-insulator and LNOI surfaces were solvent cleaned, treated with Piranha (9:1 $H_2SO_4$:$H_2O_2$), activated by a 5 min $O_2$ plasma, bonded, and annealed at 100 °C for up to 24 h. The Si handle and buried oxide were then removed by dry etching ($SF_6$/$C_4F_8$) and wet etching (6:1 buffered oxide etch). Nanoantennas were patterned by electron-beam lithography (Raith EBPG 5200+) in ma-N 2400 resist. The pattern was transferred into the Si by HBr/$Cl_2$ etching and into the LN by Ar ion milling (180 nm deep, leaving a 120 nm residual film), and the residual mask was removed in Piranha. Electrodes (2 nm Ti / 68 nm Ag) were deposited by electron-beam evaporation and lift-off (Supplementary Section 11).

### 4.3 Electro-optic characterization

Optical performance was evaluated using a custom-built reflection microscope. Static resonant spectra and active electro-optic modulation were characterized using a wavelength tunable continuous wave laser. Reflected light was collected with a 10× near infrared objective and detected using an InGaAs photodetector. For beam steering and diffraction measurements, a second InGaAs photodetector placed along the diffracted beam path collected the steered order (Fig. S13).

To extract the quality factor, the experimental and simulated spectral features were fit to a Fano–Lorentzian lineshape that describes a resonance with center frequency $f_0$ and linewidth 2γ. The Q-factor is then taken to be $Q = f_0/2\gamma$.

Electrical excitation was delivered through high-frequency microwave probes. The frequency response was recorded using high-bandwidth photodetectors (up to 1.6 GHz) connected to an electrical spectrum analyzer and oscilloscope.

We define two figures of merit for the modulated signal. The modulation efficiency is η=ΔP/Pinc, where ΔP is the peak-to-peak optical power change at the second harmonic of the drive frequency and Pinc is the total incident power. The modulation depth is m=ΔP/P0, the same power change normalized to the device's own baseline reflected power at zero bias.

**Acknowledgments.** This work was supported by the Office of Naval Research Multidisciplinary University Research Initiative (MURI) program (N00014-23-1-2567) and by the U.S. Department of Energy, Office of Science, National Quantum Information Science Research Centers, Q-NEXT (DE-AC02-76SF00515). The authors acknowledge the facilities and scientific and technical assistance provided by the Stanford Nanofabrication Facility (SNF) and the Stanford Nano Shared Facilities (SNSF).

**Competing interests.** The authors declare no competing interests.

**Data availability.** Data underlying the results presented in this paper are not publicly available at this time but may be obtained from the authors upon reasonable request.

**Supplementary Information.** Supplementary Information accompanies this paper.